\documentclass[aps,prl,onecolumn,doublespace,groupedaddress]{revtex4}
\usepackage{graphicx}
\usepackage{epstopdf}
\usepackage{hyperref}
\usepackage{color,soul}
\usepackage{amsmath,amsthm,amssymb,euscript,mathrsfs,epsf,epsfig}

\begin{document}

\title{Signature of a polyamorphic transition in the THz spectrum of vitreous GeO$_2$}
\author{Alessandro Cunsolo$^{1}$}
\email{acunsolo@bnl.gov}
\author{Yan Li$^{2}$}
\author{Chaminda~N.~Kodituwakku$^{1}$}
\author{Shibing~Wang$^{3}$}
\author{Daniele~Antonangeli$^{4}$}
\author{Filippo~Bencivenga$^{5}$}
\author{Andrea~Battistoni$^{5,6}$}
\author{Roberto~Verbeni$^{7}$}
\author{Satoshi~Tsutsui$^{8}$}
\author{Alfred~Q.~R.~Baron$^{8,9}$}
\author{Ho-Kwang~Mao$^{10,11}$}
\author{Dima~Bolmatov$^{1}$}
\author{Yong~Q.~Cai$^{1}$}
\affiliation{$^{1*}$  National Synchrotron Light Source II, Brookhaven National Laboratory, Upton, NY 11973,
USA}
\affiliation{$^{2}$ American Physical Society, 1 Research
  Road, Ridge, New York 11961, United States}
\affiliation{$^{3}$ Department of Geological and Environmental Sciences, Stanford University, Stanford, CA 94305, USA}
\affiliation{$^{4}$ Institut de Min\'{e}ralogie, de Physique des Mat\'{e}riaux et de Cosmochimie, UMR CNRS 7590, Sorbonne Universit\'{e}s - UPMC, Mus\'{e}um National d'Historie Naturelle, IRD Unit\'{e} 206, 75252 Paris, France}
\affiliation{$^{5}$ Sincrotrone Trieste, S.S. 14 km 163,5 in AREA Science Park 34012 Basovizza, Trieste, Italy}
\affiliation{$^{6}$ Dipartimento di Fisica, Università degli Studi di Trieste, I-34127 Trieste, Italy}
\affiliation{$^{7}$European Synchrotron Radiation Facility (ESRF), 71 avenue des Martyrs, 38043 Grenoble, France}
\affiliation{$^{8}$ Japan Synchrotron Radiation Research Institute, SPring-8,
Sayo, Hyogo 679-5198, Japan.}
\affiliation{$^{9}$SPring-8/RIKEN, Hyogo 679-5148, Japan}
\affiliation{$^{10}$ Geophysical Laboratory Carnegie Institution of Washington, 5251 Broad Branch Road, NW, Washington, DC 20015, USA}
\affiliation{$^{11}$ Center for High Pressure Science $\&$ Technology Advanced Research, Pudong, Shanghai, 201203, China}

\begin{abstract}
The THz spectrum of density fluctuations, $S(Q, \omega)$, of vitreous GeO$_2$ at ambient temperature was measured by inelastic x-ray scattering from ambient pressure up to pressures well beyond that of the known $\alpha$-quartz to rutile polyamorphic (PA) transition. We observe significant differences in the spectral shape measured below and above the PA transition, in particular, in the 30-80 meV range. Guided by first-principle lattice dynamics calculations, we interpret the changes in the phonon dispersion as the evolution from a quartz-like to a rutile-like coordination. Notably, such a crossover is
accompanied by a cusp-like behavior in the pressure dependence of the elastic response
of the system. Overall, the presented results highlight the complex fingerprint of PA phenomena on the high-frequency phonon dispersion.\\
* acunsolo@bnl.gov
\end{abstract}

\maketitle

\section*{Introduction}
Pressure ($P$) or temperature ($T$) induced modifications in crystal structures and associated effects on the lattice dynamics are commonly observed and reasonably well understood. On the contrary, transformations in amorphous systems between distinct aggregates having different local structure and density are more elusive. These polyamorphic (PA) transitions are often difficult to observe, since hampered by several concomitant factors. For instance, when the density is the order parameter, extreme thermodynamic conditions are required
to significantly alter this variable due to the low compressibility of amorphous, non-gaseous, systems. Furthermore, PA phenomena often happen in metastable thermodynamic regions, where they are overshadowed by competing effects, such as glass transition or crystal nucleation.

On a general ground, the best candidates to observe PA transitions are systems with an intrinsically open, often tetrahedral, local structure. In fact, the large free volume available in tetrahedral arrangements can in principle allow structural modifications even at moderate thermodynamic conditions. This was demonstrated to be the case in water \cite{Mishima}, liquid silicon \cite{Silicon}, germanium\cite{Germanium}, and phosphorus \cite{Phosphorus}, as well as in amorphous SiO$_2$ \cite{Silica} and GeO$_2$ \cite{Germania} (see Ref. [\onlinecite{Brazhkin}] for a review on the topic). In spite of a thorough experimental scrutiny, some general aspects of PA transitions are still obscure,
including the possible influence on the propagation of collective excitations. This can be particularly relevant at mesoscopic ($\sim$ nm) length-scales, where the dynamics is known to be strongly coupled with local atomic arrangements.

Inelastic neutron (INS) and x-ray (IXS) scattering are two classic experimental techniques commonly used to probe atomic and lattice motions; however severe technical difficulties hinder the observation of PA transitions.  A major one relates to the fact that modifications in the local order involving, e.g. the coordination number, usually disappear when the sample is recovered to ambient conditions \cite{nota}. This imposes \emph{in situ} high-pressure experiments on very small samples for direct observation of PA transitions. This, in most practical situations, rules out the possibility of using INS and often causes problems of spectral background in IXS measurements. Although water \cite{water_IXS} and silica \cite{silica_IXS} are the two polyamorphic materials that have been most extensively investigated by inelastic spectroscopies, no signature of PA transitions has been reported in the THz spectrum in either case, mainly owing to two different reasons: in silica PA phenomena happen at pressures still prohibitively high for scattering measurements, while in water the PA transition is expected to take place in a deeply supercooled region, representing a sort of \emph{no man's land} in the thermodynamic plane \cite{Poole}.

Compared to SiO$_2$, its structural analogous GeO$_2$ has proven to be a better candidate for IXS investigations of PA phenomena due to both the larger tetrahedral cell, which shifts the onset of PA transitions to lower $P$'s, and the higher electronic number -- and consequently shorter x-ray absorption length -- which substantially enhances the IXS signal from the small-sized sample suited for the use of Diamond Anvil Cells (DAC) \cite{BenAnt_PRB}. Accordingly, we have studied the pressure-dependent  spectrum of density fluctuation, $S(Q,\omega)$, of vitreous (v-)GeO$_2$ at ambient temperature by \emph{in situ} IXS measurements from ambient $P$ up to 26 GPa (see Methods for further details). This $P$ range has been chosen to well track the $P$-dependence below and above 9 GPa, pressure around which a sudden jump of the bond distance is reported and commonly ascribed to a transition from a tetrahedral to an octahedral local structure \cite{Germania,Smith}, or, in other terms, from an $\alpha$-quartz-like to a rutile-like local lattice organization. In a more recent x-ray diffraction work \cite{Drewitt} important structural changes have been observed to continuously occur for pressure spanning the 5-8.6 $GPa$ range, as later confirmed by oxygen K-edge IXS measurements \cite{Lelong}. Furthermore, previous studies based on classical molecular dynamics \cite{Peralta} predicted a main structural change from tetrahedra  to octahedra arrangement at 3-7 GPa, slightly lower than observed here and reported in previous works.

In addition to diffraction and absorption measurements \cite{Micolaut}, signatures of a PA transition in v-GeO$_2$ have been sought for by investigating the vibrational behavior by Raman and infrared techniques \cite{Micolaut,Galeener}, which provided evidences of possible dynamic counterparts of the aforementioned PA transition. Combining experimental results with first-principle density functional theory (DFT) calculations, we aim at detecting and explaining the signature of the PA crossover in the THZ spectrum of v-Ge$O_2$.

\section*{Results}

Typical IXS spectra measured at ambient pressure and at $P =$ 26 GPa, i.e. respectively well below and above the PA transition, are shown in Fig. \ref{Fig1} along with the experimental resolution function. We recall here that PA transition of GeO$_2$ is here identified with the large transformation of the average Ge-O bond distance, $d_{O-Ge}$ in the $ 5 \lesssim P \lesssim 9 \,GPa$ range(see Ref. \cite{Germania} and \cite{Smith}). Although, as mentioned, such a crossover was also either observed as a sharp, yet continuous, P-increase \cite{Drewitt,Lelong} or to occur at slightly lower P values, it is unanimously found that it takes place for P $<$ 10 GPa.
A clear excess of scattering intensity with respect to the resolution profile can be observed in all spectra, indicating the presence of an appreciable contribution of collective excitations to the experimental spectra in all the probed ($Q$,$P$)-values. Such inelastic spectral wings turn into a ``double shoulder'' lineshape at higher $Q$'s for both $P$ values, hence suggesting the presence of two distinct collective modes. In order to determine the $Q$-dispersion relations of such modes and gain insights into their correlations to structural changes within the probed $P$-range, we performed a best-fit line-shape analysis to determine the characteristic frequencies of the high ($\Omega_\mathrm{HF}$) and low frequency ($\Omega_\mathrm{LF}$) excitations (see Methods for further details).

The dispersion curves obtained as the best-fit values of the $\Omega_\mathrm{LH}$ and $\Omega_\mathrm{HF}$ parameters in Eq. \ref{DHO_2} are displayed in Fig. \ref{Fig2}.
It may be noticed that at the lowest  $Q$s some $\Omega_\mathrm{LH}$ are not reported in the plot; this owes to the corresponding vanishing intensity of the low frequency peak in the spectrum.\\

Even more striking is the transformation of the $P$-evolution of the generalized sound velocities $c_s$ (Fig. 3) extracted from the low Q ($\leq 10 nm^{-1}$) slope of the dispersions curves in Fig. \ref{Fig2}. The fact that IXS probes sound propagation at THZ frequencies and $nm$ distances makes IXS measurements more  sensitive to local molecular arrangements and interactions than ultrasound or Brillouin light scattering (BLS) techniques. Derived velocities can thus differ from ultrasonic and hypersonic determinations (see for instance discussion in Ref. \cite{BenAnt_PRB}), bringing complementary information. To further stress the distinctive elastic behavior of the low-pressure and high-pressure polyamorph, the inset of Fig. \ref{Fig3} shows the value of the longitudinal modulus $ M = \rho c_s$ - with $\rho$ being the mass density, here derived from from Ref. \cite{Smith}.

Figs. \ref{Fig4} and \ref{Fig5} compare, respectively, the lowest $P$ ($P$ = 0 GPa) and highest $P$ ($P$ = 26 GPa) dispersions shown in Fig. \ref{Fig2} with the phonon dispersions computed for crystalline $\alpha$-quartz GeO$_2$ and rutile GeO$_2$ at the corresponding $P$s along high symmetry directions; here the thickness of the dispersion curves is proportional to the weight of the corresponding contributions to the $S(Q,\omega)$ (see Methods for further details). The green lines in Figs. \ref{Fig4} and \ref{Fig5} denote the $(\omega,Q)$-range dominated by the phonon modes of diamond. Within this region the very intense scattering contribution from the diamond anvils hamper the detection of the signal from the sample. Right panels in Figs. \ref{Fig4} and \ref{Fig5} show the computed total vibrational density of state (v-DOS). The v-DOS measured  by Raman scattering at $P$ = 0 GPa \cite{Deshamps} and 32 GPa \cite{Durben} are also included for comparison \cite{note}
in Figs. \ref{Fig4} and \ref{Fig5}.

\section*{Discussion}
The quality of the data analysis can be readily judged from the overall good agreement between measured and best-fit line-shapes in Fig. \ref{Fig1}, where the spectral contributions of the low and high frequency modes to the total scattering intensity are also shown. By comparing inelastic modes in the spectra measured at $P$ = 0 GPa and $P$ = 26 GPa one immediately observes that both inelastic shift and relative intensity of the high frequency mode increase dramatically with pressure.

A substantial difference between low- and high-$P$ sound dispersions readily emerges from Fig. \ref{Fig2}: $\Omega_\mathrm{LF}$ and $\Omega_\mathrm{HF}$ at low $P$s (left panel) span over lower energy values and exhibit a moderate $Q$-dependence. In contrast, the high-$P$ dispersions (right panel) are featured by a strongly $Q$-dependent high-$\omega$  phonon branch and by a mildly $Q$-dependent low-$\omega$  branch. Such features become even more striking when considering that $P$ effects are relatively moderate within each subset of data. This result is in good agreement/compatible with the expected transition value of 9 GPa.\\
Further transformations of the acoustic properties of the sample across the PA crossover clearly emerge in Fig. \ref{Fig3} through the cusp-like $P$ dependence of $c_s$.\\
The drastic change is emphasized by the different slopes of the straight lines best fitting all reported $c_s$ data either below or above the crossover (dashed lines). These lines intercept at a P value ( $\approx$ 8.7 GPa) close to the known PA crossover pressure. The lower high $P$ slope clearly indicates a higher resistance of the sample to $P$-induced modifications of elastic properties.  In a similar manner, in recent BLS
works \cite{Murakami,Huang} changes in the slope of the P-evolution of sound velocity and/or elastic moduli were interpreted as signatures of PA phenomena; more specifically, they have been correlated to changes in the coordination number. Here
the cusp-like behavior exhibited by $c_s$ data seems more pronounced. This is the likely consequence of the mentioned rough matching between probed distances and first neighboring molecules'.

As the crystalline $\alpha$-quartz and rutile structures are known to approximate first neighbor Ge-O bonding arrangements of v-GeO$_2$ for $P$ below and above the PA crossover \cite{Germania}, we use DFT results as an interpretative guidance for the observed $P$-evolution of the collective dynamics measured on glasses. Bearing in mind the different sample nature (crystalline vs glassy), experimental and computational results shown in Figs. \ref{Fig4} and \ref{Fig5} are overall consistent. In particular, it is worth noticing that both computed dispersions and v-DOS exhibit distinctive features, markedly different between the $\alpha$-quartz and rutile phases. Indeed, dispersion curves of $\alpha$-quartz clearly span over lower energies, and optical phonon modes exhibit a classical relatively mild $Q$ dependence. Accordingly, the v-DOS of this structure is characterized by the expected parabolic low-frequency trend, followed by van Hove singularities of comparable intensity and, finally, by a clear phonon gap in the 45-55 meV interval (see inset in Fig. \ref{Fig3}).
 In contrast, for the the rutile lattice arrangement, the main contribution to $S(Q,\omega)$ comes from highly dispersive high-$\omega$ phonon branches. The v-DOS is dominated by a large van Hove singularity at about 45 meV and presents no evidence of phonon gaps below 60 meV. As a consequence, within the reported 0-60 meV range, the v-DOS of rutile spans comparatively higher $\omega$'s than that of $\alpha$-quartz. In view of these qualitative differences, the overall agreement of experimental data collected at pressures below 9 GPa with calculations for the  $\alpha$-quartz structure (Fig. \ref{Fig4}) and of the experimental data collected at pressure above 9 GPa with the calculations for the rutile crystal structure (Fig. \ref{Fig5}) provide strong indications that the PA quartz-like to rutile-like transition in the local structure of v-GeO$_2$ \cite{Germania} has a well-defined counterpart in the phonon dispersion behavior.

A closer inspection of Fig. \ref{Fig5} reveals few additional interesting aspects. The IXS measurements unmistakably show that, at high $P$, the $S(Q,\omega)$ of v-GeO$_2$ is dominated by two modes with very different energy, one highly $Q$-dispersive, and a second weakly $Q$-dispersive. The presence of the latter was particularly evident in the Spring-8 data (P= 13 GPa) most likely because of the better contrast in the low-frequency spectral range (see Methods). The comparison with the computed dispersion curves of the rutile crystal allows us to tentatively associate these two branches to longitudinal (LA) and transverse acoustic (TA) modes, respectively.
In order to facilitate the comparison between $\Omega_{LF}$ values and the TA modes in the crystal, in Fig. \ref{Fig4} the lowest frequency TA branches of $\alpha$-quartz are also indicated as dashed lines in the three crystalline paths considered. However, the spectral contribution of these branches was found negligible in the crystal.
Whereas for the crystal phase TA contributions to $S(Q,\omega)$ are forbidden within the first Brillouin zone, they may become sizable in the glassy phase. This directly relates to the absence of long-rage translational symmetries, or equivalently well-defined Brillouin zones. "Pure" symmetric TA modes can be "contaminated" and acquire a mixed longitudinal and transverse character. Such mode-mixing is often referred to as the longitudinal-transverse (L-T) coupling \cite{Sampoli} and is the physical rationale behind the appearance of a shear mode in the $S(Q,\omega)$, which primarily couples with longitudinal movements only. It is worth noticing that a similar L-T coupling has been reported in various systems sharing a tetrahedral molecular arrangements, such as water \cite{water}, GeSe$_2$ \cite{GeSe2} and v-GeO$_2$ as well \cite{Bove}. It can thus be envisaged that the open and highly directional nature of such arrangement fosters the onset of an L-T coupling.

Furthermore, the comparison between left and right panels in Fig. \ref{Fig5} indicates that the value of $\Omega_\mathrm{LF}$ well corresponds to the low frequency excess of the v-DOS in the glass as compared to the crystal.
This intensity excess relates to almost universal feature of glasses essentially amounting in a peak in the reduced density of states v-GeO$_2/E^2$, customarily referred to as Boson peak (BP) \cite{nota2}.
The equivalence between the BP of a glass and the Van Hove singularity of the TA branch of the corresponding crystal has been recently demonstrated to be a sound hypothesis \cite{Chumakov}. We observe that the intensity ratio between TA and LA modes decreases upon increasing the pressure consistently with the observed pressure trend of the BP \cite{Niss}.

\section*{Conclusion}

In summary, we investigated the evolution of the THz spectrum of density fluctuations in v-GeO$_2$ as a function of pressure. We observed a clear transformation in the $Q$-dispersion behavior upon crossing the known polyamorphic transition occurring in the glassy phase at $\approx$ 9 GPa. Supported by DFT calculations, we interpreted this as the abrupt evolution from a quartz-like to a rutile-like behavior, concluding that the collective dynamics of v-GeO$_2$ in the THz range is strongly sensitive to the undergoing changes in the local structure. This is clearly indicated by a cusp-like behavior of the pressure dependence of generalized sound velocity and longitudinal modulus across the PA transition. Both these trends suggest that the high-P polyamorphic phase is characterized by a higher resistance to pressure induced modification of elastic properties, likely due to the more packed first neighbor arrangement.
Furthermore, presented data indicates that the inherent disorder characteristics of the glassy phase seems to foster the visibility of a low frequency transverse modes, which, especially at low pressures, is evident at high $Q$'s but it is still appreciable within the first pseudo-Brillouin zone. This mode in the glass is here found to be possibly related to the low frequency excess intensity in the vibrational density of state.

We finally remark that, when compared to more traditional structural measurements, investigations of the THz dynamics as the one presented in this work provide a complementary insight onto the PA transition linking it to transformations of elastic properties.

More in general, the results presented in this work pave the way toward future investigations of the dynamic aspects of polyamorphism phenomena by using THz probes of phonon-like modes.

\section*{Methods}
\subsection{IXS measurements and data analysis}

Two independent IXS experiments were carried out on two different IXS beamlines: beamline BL35XU \cite{BL35} at SPring-8 (Hyogo, Japan) and beamline ID28 \cite{ID28} at the European Synchrotron Radiation Facility (ESRF, Grenoble, France). These two triple-axis spectrometers have the same working principle, based upon high order reflections from nearly perfect Si crystals. The instruments were operated at the Si$(9,9,9)$ configuration for both monochromator and analyzer crystals, corresponding to an incident energy of 17.947 KeV and an overall resolution bandwidth of 3.0 meV (FWHM). The profile of the instrumental resolution function, $R(\omega)$ was estimated through the measurement of the spectral line-shape of an essentially elastic scatterer, specifically a cryogenically cooled sample of perspex at the $Q$ position of its first sharp diffraction peak (10 nm$^{-1}$). The spectrometer was rotated to probe within a single energy scan the 4.5-7.8 nm$^{-1}$ $Q$-range (BL35XU measurements) or the 5.4-13.8 nm$^{-1}$ $Q$ range (ID28 measurements). The focal spot of the beam at sample position was 35$\times$15 $\mu m^2$ and 30$\times$60 $\mu m^2$(horizontal x vertical FWHM) for measurements at BL35XU and ID28, respectively.

The v-GeO$_2$ sample was synthesized as described in a previous work~\cite{Mei_PRB_2010} (BL35XU experiments) or purchased (Sigma-Aldrich; ID28 experiments). In both cases, the samples were loaded in DAC without any pressure transmitting medium to maximize the scattering volume for high-pressure measurements. Samples used for BL35XU experiments were conditioned in Mao type DAC, equipped with tungsten gaskets, while samples used for ESRF experiments were conditioned in membrane driven Le Toullec type DAC, equipped with rhenium gasket. We stress that the metallic gasket, beside radially containing the sample, also acts as cleaning pinhole in the x-ray path. To further minimize, spurious quasi-elastic scattering contributions, DACs were placed in a specifically designed vacuum chamber equipped with motorized exit slits. We indifferently used 350 or 300 $\mu m$ flat culets diamond to cover the pressure range of interest.
A small chip of ruby placed in the sample chamber served as pressure gauge \cite{Ruby}. In order to measure the spectral lineshape with the needed statistical accuracy, we performed energy transfer scans typically lasting 18-24 hours. Prior to each spectral acquisition we measured the $Q$-dependent elastic scattering intensity $I(Q,\omega=0)$, systematically confirming the vitreous nature of the investigated phase.

The pressures probed in the SPing-8 experiment were 3.7, 13 and 20 GPa, while in the ESRF experiment we collected data at room pressure and 26 GPa, in both cases the sample was at ambient temperature. SPring-8 measurements were performed with a better spectral contrast in the low-frequency spectral region, likely due to the lower quasi-elastic scattering from the diamonds DAC windows, while in spectra collected at ESRF had substantially higher count-rate and, consequently, a better statistical accuracy.

\begin{figure}[ht]
\begin{center}
\includegraphics[height =3.5 in]{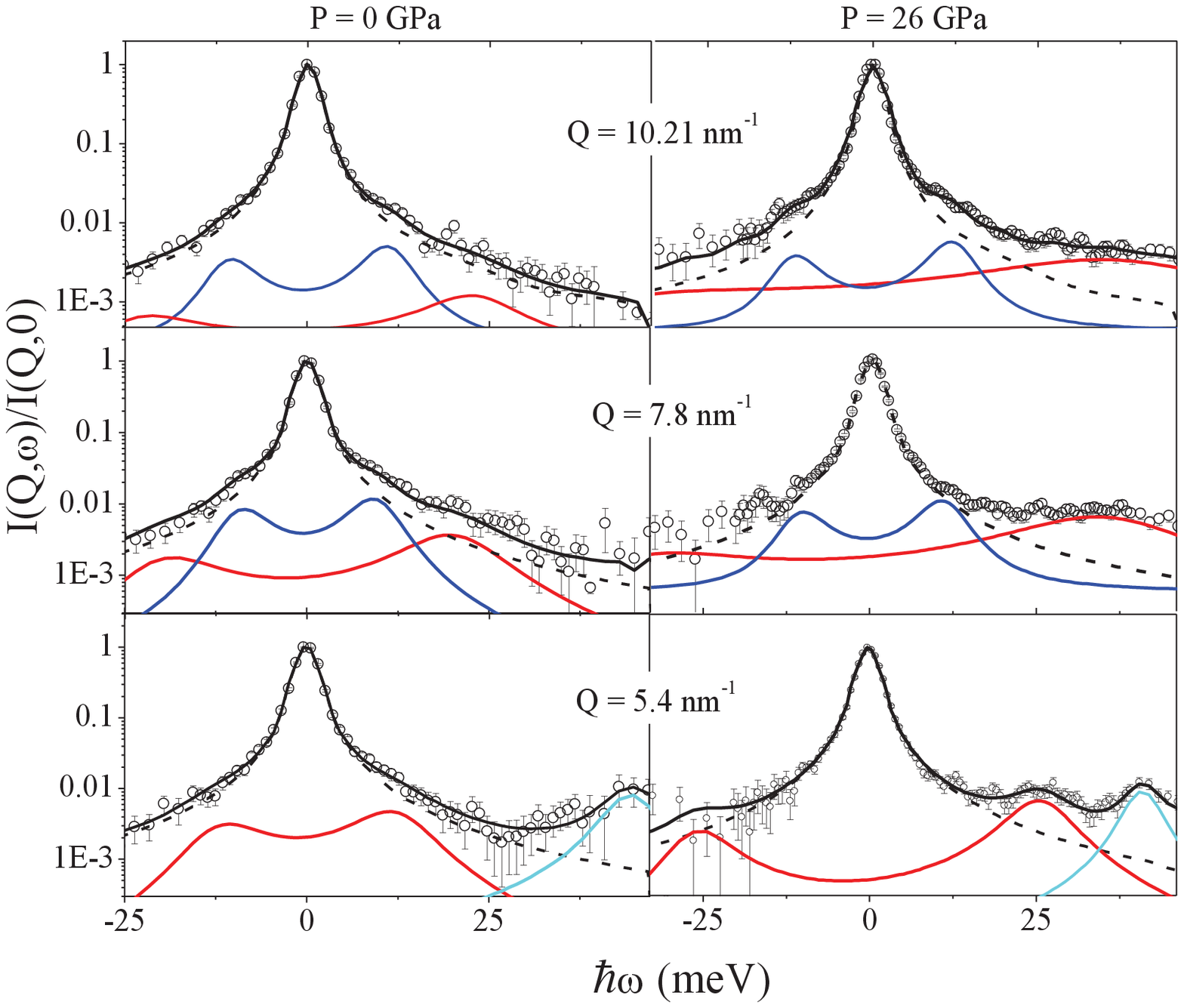}
\caption{{\bf IXS spectra below and above the PA crossover}. Representative IXS spectra of v-GeO$_2$ measured at low and high pressures for selected $Q$ values (open circles). The thick black line corresponds to the best-fit model lineshape in Eq. \ref{DHO_2}, with its low frequency (blue line) and high frequency (red line) DHO components. The dashed and cyan lines represent respectively the resolution function and a DHO profile accounting for the transverse mode of the diamond anvil cell.}
\label{Fig1}
\end{center}
\end{figure}

\begin{figure}[ht]
\begin{center}
\includegraphics[height =3.5 in]{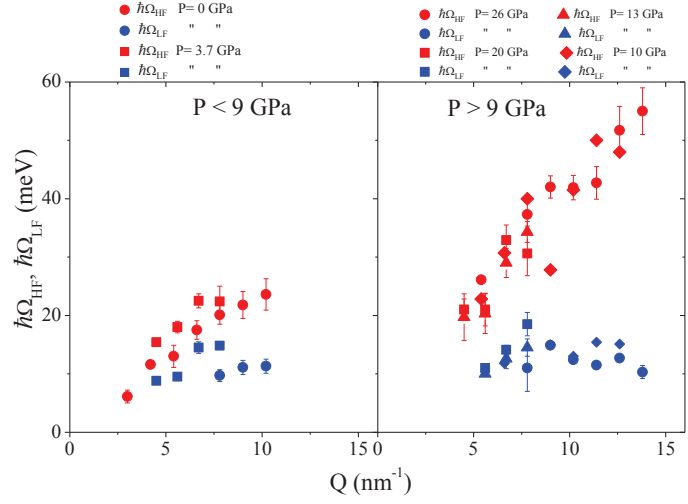}
\caption{{\bf Dispersion curves below and above the PA crossover}. Best-fit dispersion of $\Omega_\mathrm{LF}$ (blue symbols) and $\Omega_\mathrm{HF}$ (red symbols) as measured at the indicated pressures below (left panel) and above (right panel) the PA transition.}
\label{Fig2}
\end{center}
\end{figure}

The best-fit line-shape modeling of IXS spectra is based on a sum of two Damped Harmonic Oscillator (DHO) functions plus an essentially elastic central peak. Overall the used profile reads as

\begin{eqnarray}
\frac{S(Q,\omega)}{S(Q)}=f(\omega)\left[A\delta(\omega)+I_\mathrm{LF}\frac{\left(2\Omega_\mathrm{LF}\Gamma_\mathrm{LF}\right)^2}{(\omega^2-
\Omega_\mathrm{LF}^2)^2+
4\Gamma_\mathrm{LF}^2\omega^2}+I_\mathrm{HF}\frac{\left(2\Omega_\mathrm{HF}\Gamma_\mathrm{HF}\right)^2}
{(\omega^2-\Omega_\mathrm{HF}^2)^2+4\Gamma_\mathrm{HF}^2\omega^2}\right],
\label{DHO_2}
\end{eqnarray}

\noindent where $f(\omega)=\hbar\omega/k_BT[1-\exp(-\hbar\omega/k_B T)]$ is the detailed balance factor accounting for the frequency-dependent statistical population of the states, with $k_B$ being the Boltzmann constant and $\hbar=h/2\pi$ with $h$ being the Planck constant. The parameters $\Omega_\mathrm{LF} (\Omega_\mathrm{HF})$ and $\Gamma_\mathrm{LF} (\Gamma_\mathrm{HF})$ represent the shift and the width the low (high) frequency inelastic mode, respectively, while $A$, $I_\mathrm{LF}$ and $I_\mathrm{HF}$ are frequency independent scaling factors. A $Q$-dependence is assumed implicitly for all parameters. The double DHO profile of Eq. \ref{DHO_2} was used to fit the measured spectral line-shape, even though at the lowest $Q$s best-fit values of $I_\mathrm{LF}$ were negligible. Additional DHO terms were added to account for the inelastic scattering from the phonon modes of diamond. Due to the very high sound speed of diamond, such well resolved spectral features locate in the high-frequency side of the spectra and disperse out of the probed window at high $Q$'s (see Fig. \ref{Fig1}).

Ultimately, the best fit of measured spectra (see Fig. \ref{Fig1}) was achieved through the minimization a $\chi^2$ variable defined as the normalized distance between the experimental line shape and the following model profile:

\begin{equation}
I(Q,\omega) = K S(Q,\omega)\otimes R(\omega) + B,
\label{final}
\end{equation}

The symbol "$\otimes$" represents the convolution operator, while $K$ and $B$ are two frequency-independent constants representing, respectively, an overall intensity factor and a flat background, which also includes the electronic noise of the detectors ($<$ 1 mHz).

\subsection{Numerical computations}

To get further insights into the dynamic response of the sample, we complemented our experimental investigation with a numerical study. DFT calculations within the local density approximations were carried out using the ABINIT package~\cite{Gonze} and norm-conserving pseudopotentials in the Troullier-Martins scheme. The crystal structures of bulk GeO$_2$ were optimized using a kinetic energy cutoff of 58 Hatree, and the first Brillouin zone were sampled using a $4\times 4\times 4$ and $6\times 6\times 6$ $Q$-grid for the $\alpha$-quartz and rutile structures, respectively. Phonon dispersion curves were interpolated over a $4\times 4\times 4$ and $3\times 3\times 3$ $Q$-grid for $\alpha$-quartz and rutile structures of GeO$_2$, respectively, using the density-functional perturbation theory scheme described in ~\cite{Lee}.  Nonanalytic corrections due to the long-ranged anisotropic dipole-dipole interactions were applied to the dynamic matrix. The computed zone-center phonon frequencies at $P=0$ GPa for $\alpha$-quartz and rutile structures are within 0.6 meV of the experimental values~\cite{Madon,Sharma} and previous DFT results~\cite{Hermet}. The weight of the individual contributions to $S(Q,\omega)$ at a discrete ${\bf Q}_i$ on the $j$-th phonon branch (represented in Figs. \ref{Fig3} and \ref{Fig4} as the thickness of the corresponding line after multiplying with $\omega({\bf Q}_i)$) is denoted as $S_{j}({{\bf Q_i}})$:

\begin{eqnarray}
S(Q,\omega)&=&\sum_{i,j}S_{j}({{\bf Q}_i})\delta(\omega-\omega_j({\bf Q}_i)),
\end{eqnarray}

with:

\begin{eqnarray}
S_{j}({{\bf Q}_i})&=&\frac{1+n_j({\bf Q}_i)}{2\omega_j({\bf
    Q}_i)}\bigg|\sum_d\frac{f_d({\bf Q}_i)}{\sqrt{M_d}} {\bf Q}_i\cdot
\sigma_d^j({\bf Q}_i) \mathrm{exp}^{\imath{\bf Q}_i\cdot{\bf
    d}}\bigg|^2 \nonumber\\
\end{eqnarray}

\noindent where $f_d({\bf Q})$ is the x-ray form factor and $\sigma_d^j({\bf Q})$ is the projection on to atom $d$ of eigenvector in the $j$-th branch. Effects due to the Debye-Waller factor were ignored since it induces a negligible $Q$ dependence within the probed momentum transfer interval.

\section*{Figure captions}
\begin{figure}[ht]
\begin{center}
\includegraphics[height= 3.5 in ]{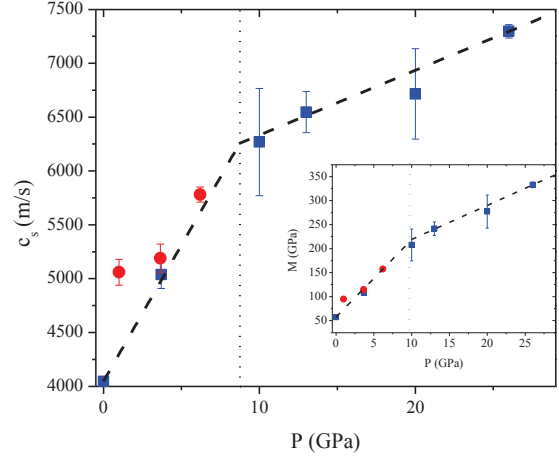}
 \caption{{\bf P-dependence of the sound velocity across the PA crossover.} The sound velocities extracted from the low $Q$ slope of the dispersion curves in Fig. \ref{Fig2} (see text) are reported (blue squares) as a function of pressure. The dashed lines are the outcome of a linear fit to data below and  data below and above the PA crossover. The pressure value at which the two straight lines intersect ($\sim8.7 GPa$) is highlighted by a vertical dotted line. The red dots are the velocities extracted from Ref. \cite{BenAnt_PRB}. The inset shows the corresponding longitudinal modulus M, as derived using density data in Ref. \cite {Smith}.}
\end{center}
\end{figure}[ht]
\begin{figure}
\begin{center}
\includegraphics[height = 3.5 in]{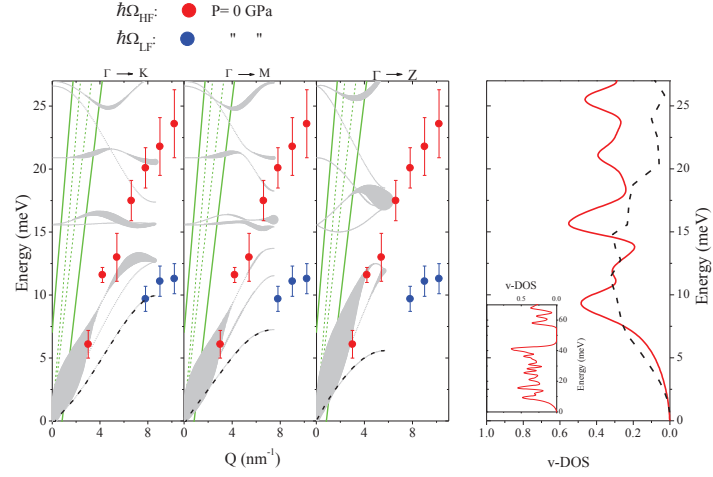}
 \caption{{\bf Comparison between experimental and theoretical results below the PA crossover}. Left: Dispersion curves along high-symmetry directions of $\alpha$-quartz GeO$_2$ computed at $P=$ 0 GPa (black lines) and $P=$ 0 GPa values of $\Omega_\mathrm{LF}$ (blue symbols) and $\Omega_\mathrm{HF}$ (red symbols) measured in v-GeO$_2$. The thickness of computed dispersions is proportional to the corresponding $S(Q,\omega)$ contribution (see text). In the three plots the lowest frequency TA branch is also reported for comparison as dashed line. The green curves represent the boundary of the zone dominated by the diamond's phonons. Right: computed v-DOS (red line) of $\alpha$-quartz GeO$_2$ at $P=0$ GPa, compared with Raman scattering measurements in v-GeO$_2$ at $P=0$ GPa \cite{Deshamps} (dashed line), after re-scaling for an arbitrary factor. The inset displays the calculated v-DOS in an extended energy range.}
\label{Fig4}
\end{center}
\end{figure}
\begin{figure}
\begin{center}
\includegraphics[height=3.5 in ]{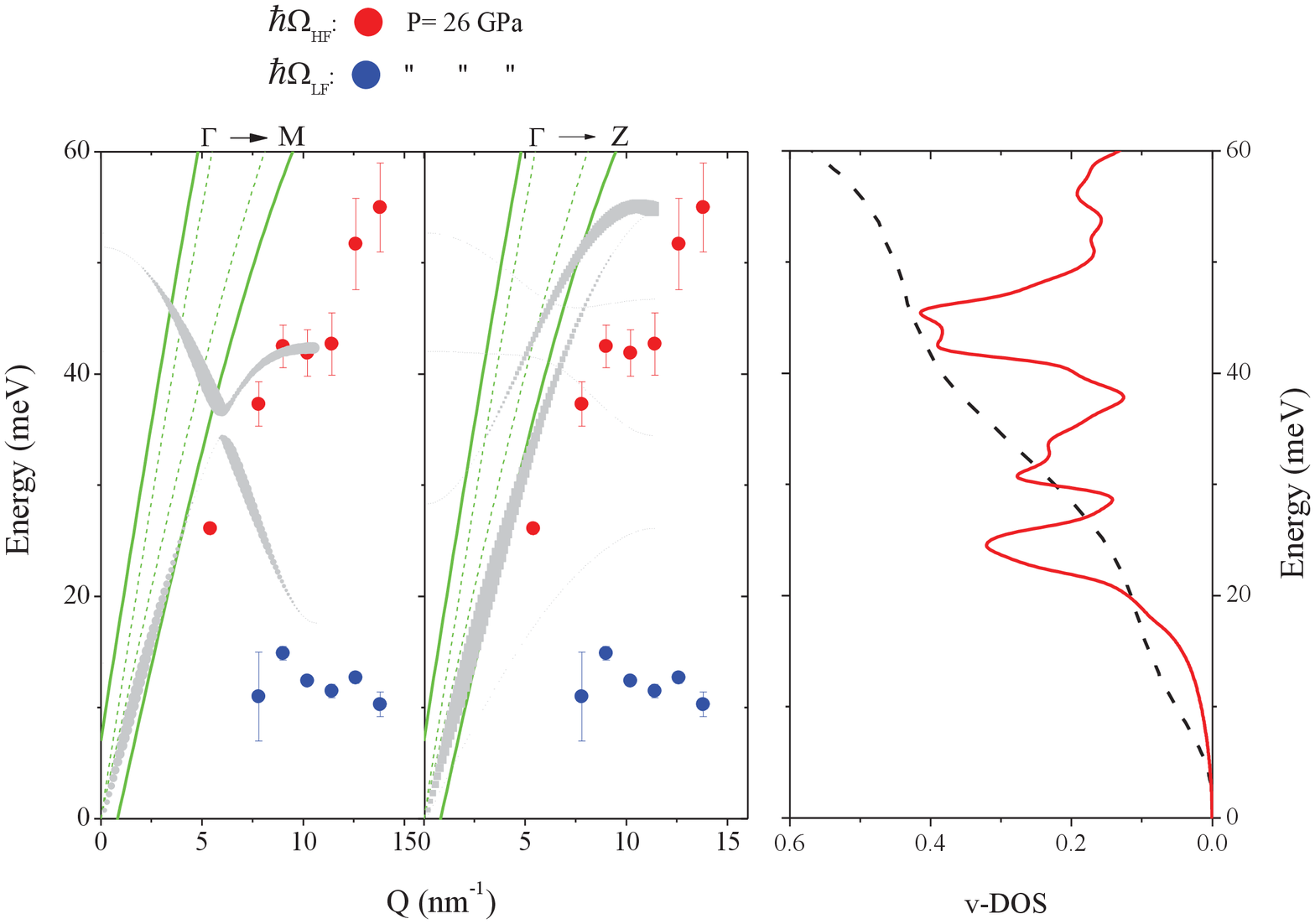}
 \caption{{\bf Comparison between experimental and theoretical results above the PA crossover}. Same as Fig. \ref{Fig3} for the rutile GeO$_2$ at $P$ = 26 GPa. The DFT computed dispersions are compared to the $P=$ 26 GPa values of $\Omega_\mathrm{LF}$ (blue symbols) and $\Omega_\mathrm{HF}$ (red symbols) measured in v-GeO$_2$ at $P=26$ GPa. The corresponding v-DOS in the right plot is compared to the Raman scattering profile taken from Ref. [\onlinecite{Durben}] at $P$ = 32 GPa.}
\label{Fig5}
\end{center}
\end{figure}
\newpage

\section*{Acknowledgements}
This work was supported by U. S. Department of Energy, Office of Science, Office of Basic Energy Sciences, under Contract No. DE-AC02-06CH11357. The authors acknowledge the technical and scientific support from both Spring-8 and ESRF staff. A.C., C.N.K., D.B. and Y.Q.C.acknowledge NSLS-II project for funding the travels for the measurements. DA acknowledge financial contribution from the French National Research Agency (ANR) through Grant 2010-JCJC-604-1. Calculations were performed at the National Energy Research Scientific Computing Center. Measurements at the beamline at BL35XU of SPring-8
were carried out using beamtime granted to the 2012A1122 research proposal.

\section*{Author contributions statement}
A.C., D.B., D.A., F.B., Y.L., and Y.Q.C. designed the research. A.B., A.C., A.Q.B, C.N.K., D.A., F.B., H-K.M., R.V., S.W., S.T. and Y.Q.C. prepared the high pressure cell and performed IXS measurements. Y.L. performed the numerical simulations. F.B. carried out the data analysis and A.C. wrote the manuscript. All authors discussed the results and commented on the manuscript.

\section*{Competing financial interests}
The authors declare no competing financial interests.

\begin{thebibliography}{99}
\bibitem{nota}
The same doesn't apply to long-range "densification" effects which in glasses are usually permanent.
\bibitem{note}
The DOS was derived from the Raman measurements as prescribed by Eq. 2 of Ref. \onlinecite{Deshamps}, yet without correction for the light-vibration coupling facor.
\bibitem{nota2}
The position of the Boson peak is red-shifted with respect to the first van Hove singularity in the crystal owing to the force-constants disorder typical of glassy materials, which induces a softening of TA modes in the glassy phase.
\end{thebibliography}
\end{document}